\documentclass[conference]{IEEEtran}
\IEEEoverridecommandlockouts

\usepackage{cite}
\usepackage{amsmath,amssymb,amsfonts}
\usepackage{algorithmic}
\usepackage{graphicx}
\usepackage{textcomp}
\usepackage{xcolor}
\renewcommand{\thetable}{\arabic{table}}

\AtBeginDocument{%
  \providecommand\BibTeX{{%
    \normalfont B\kern-0.5em{\scshape i\kern-0.25em b}\kern-0.8em\TeX}}}

\begin{document}

\title{Improving the Quality of Commit Messages \\in Students' Projects}

\author{
\IEEEauthorblockN{Iris Ma}
\IEEEauthorblockA{\textit{Department of Informatics} \\
\textit{University of California, Irvine}\\
Irvine, USA \\
huaiyaom@uci.edu}
\and
\IEEEauthorblockN{Cristina V. Lopes}
\IEEEauthorblockA{\textit{Department of Informatics} \\
\textit{University of California, Irvine}\\
Irvine, USA \\
lopes@uci.edu}
}

\maketitle

\begin{abstract}
Commit messages play a crucial role in collaborative software development. They provide a clear and concise description of the changes made to the source code. However, many commit messages among students' projects lack useful information. This is a concern, as low-quality commit messages can negatively impact communication of software development and future maintenance. To address this issue, this research aims to help students write high-quality commit messages by ``nudging'' them in the right direction. We modified the GitHub Desktop application by incorporating specific requirements for commit messages, specifically ``what" and ``why" parts. To test whether this affects the quality of commit messages, we divided students from an Information Retrieval class into two groups, with one group using the modified application and the other using other interfaces. The results show that the quality of commit messages is improved in terms of informativeness, clearness, and length.

\end{abstract}


\section{Introduction}
Software engineering is a field that heavily relies on human resources. As a social activity, software projects sometimes suffer from poor communication, which can be a challenge during development and maintenance. Commit messages are an essential tool to address communication in software development, because they provide a historical record of changes made to the source code, the reasons for those changes, and any other relevant information that developers need to understand and review the code. 

Unfortunately, the quality of commit messages is often poor. Studies have found that a significant proportion of commit messages in open-source projects lacks enough information. For instance, Dyer et al. found that about 14\% of commit messages in more than 23,000 Java projects hosted on SourceForge are empty, and only 10\% of commit messages contain at least one sentence-long descriptive information~\cite{boa-infrastructure-2013}. More recently, Tian et al. found that more than 40\% of commit messages on popular open-source projects need to be improved~\cite{a-good-commit-2022}. Low-quality commit messages can lead to confusion and difficulty in understanding the changes made to the code, making it more difficult for developers work effectively in the process of developing, reverting, and maintaining the projects.

In the context of education, the quality of commit messages in student projects can vary greatly, but overall, it tends to be poor. According to~\cite{doubt-try-three-2022}, without proper training, the percentage of good quality commits was reported to range from 27\% to 38\%. Learning how to compose these messages properly should start in school. Previous research has focused on improving overall commit behavior, with Berg et al.~\cite{doubt-try-three-2022} proposing a teaching process that emphasizes committing periodically, focusing on a single task, and providing informative commit messages. The results of their approach are promising in promoting better version control and commit behavior among students. However, such process may take at least one-week long and hours of tutorial sessions. It can be resource-intensive and may not be feasible for certain classes.

For the past several years, we have been teaching a large project-based class on Information Retrieval at UC Irvine. This class is focused on understanding search engines, and there is very little time for teaching students about good software engineering practices. The students work in groups to develop a complete Web search engine. At the end of the course, the groups meet with one of the Teaching Assistants, and the members of the group are asked questions individually. Commit messages are important for three reasons: (1) they are a means for the members of the group to communicate with each other during the project development, (2)  they are a means for students to review what they did before meeting with the Teaching Assistant; and (3) they are used by the Teaching Assistants to determine the contributions of individual students. 

Because there is no time to properly teach software engineering practices in this course, we set out to explore lightweight methods that could nudge students into writing good commit messages without requiring them to read guideline documents. Tian et al.'s work~\cite{a-good-commit-2022} makes the observation that "good" commit messages typically consist of two parts: what changed and why. Inspired by this observation, we decided to test whether a slightly modified user interface of a popular version control tool, GitHub Desktop,\footnote{https://desktop.github.com/} would be enough to guide them into better commit messages. The results we collected indicate that is the case.

This paper describes the changes we made to GitHub desktop, how we designed the study, and presents the results we obtained.

\section{Study Design}

Our study was integrated in the Web search engine project in the Information Retrieval class offered at UC Irvine in the Fall of 2022. The course focuses on providing students with a basic understanding of web information retrieval and the functioning of modern search engines. As part of the class, students  complete three coding projects, either individually or in small groups of no more than four people. The assignments include tasks related to text processing, web crawling, and building an inverted index and a UI for search. There were 120+ students enrolled in this class divided into 48 groups. All groups are required to host their projects on GitHub, and they are encouraged (but not required) to use the GitHub Desktop application to commit changes. 

\begin{figure}[htbp]
\centerline{\includegraphics[width=0.6\columnwidth]{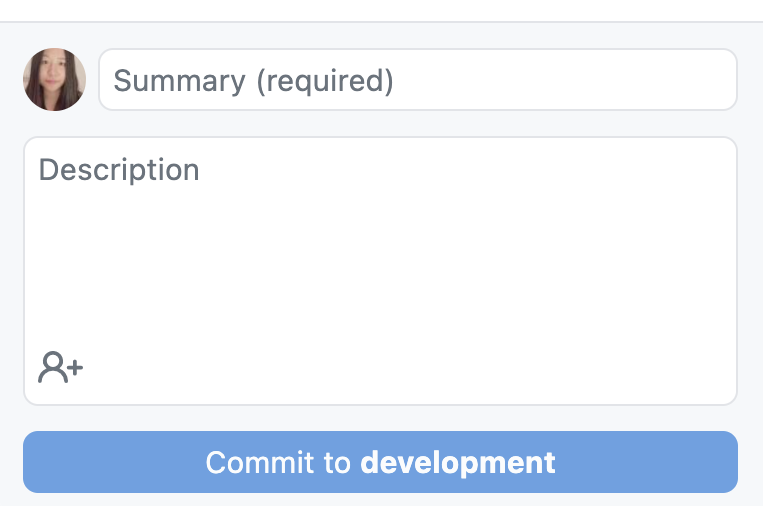}}
\caption{Original GitHub Desktop Application}
\label{original}
\end{figure}

\begin{figure}[htbp]
\centerline{\includegraphics[width=0.6\columnwidth]{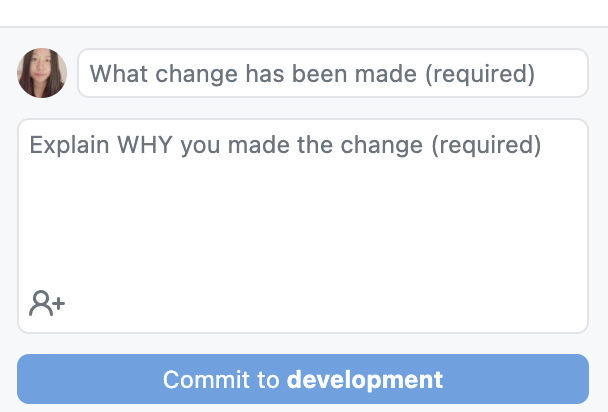}}
\caption{Modified GitHub Desktop Application}
\label{modified}
\end{figure}

The standard GitHub Desktop application (Fig.~\ref{original}) requires users to provide a summary and, optionally, a description of their changes. This follows the widely accepted form of commit messages in every source control system. For the purpose of our study, however, we modified the application in two subtle ways: first, we changed the language of those two parts to ``what'' and ``why'' (Tian et al.~\cite{a-good-commit-2022}); second, we made the second part (i.e. ``why'') required -- see Fig.~\ref{modified}.

Odd-numbered student groups were asked to use version A (the modified version of the GitHub Desktop application), while the even-numbered groups were asked to use version B (the official version). The students were encouraged, but not required, to use the appropriate version. Additionally, students were not provided with information regarding the modifications made to version A, nor were they informed the differences between versions.

Some students followed our recommendations, while other students did not; a few students used the command line interface to git rather than the GUI tool. We also did not give them any instructions about how to write commit messages, or even how to use GitHub or the GitHub Desktop application. We asked them to submit a screenshot of the tool they used along with their final submission package, so we could check. Again, some students submitted the screenshot, while others did not.

At the end of the quarter, and as part of the normal grading duties of the Teaching Assistants, we collected the commit messages from each group's GitHub repository as a factor to their individual score. Although the quality of the commit messages did not affect their score (only their number did), we compared the quality of commit messages made using the the modified version of the application against the quality of the commit messages made using other interfaces.

\section{Results}

In total, there were 1,593 commits collected. Before doing any comparisons, we filtered out the auto-generated merge-related commit messages that start with ``Merge branch,''  ``Merge pull request'' and ``Merge remote-tracking branch;'' 
Additionally, we removed commits made before the start date of using the applications, November 4, 2022. Also, some students did not provide their GitHub repository links or  screenshots for the application used, so we also excluded commits whose creation interface could not be determined. 
After these filters were applied, 801 commits remained for analysis. Table~\ref{tab:tools} shows the distribution of commits among the various interfaces used by the students. 

\begin{table}
\caption{Distribution of commit interface tools}
\begin{center}
\begin{tabular}{l c}
\hline
\textbf{\textit{Tool}}& \textbf{{\# of commits}} \\
\hline
Command Line & 105 (13.1\%)\\
\hline
Official Desktop App & 571 (71.3\%)\\
\hline
Modified Desktop App & 125  (15.6\%)\\
\hline
\end{tabular}
\label{tab:tools}
\end{center}
\end{table}

Following the method used in Tian et al.’s work~\cite{a-good-commit-2022}, we manually classified these messages into four categories depending on whether their contents included ``what" and ``why" parts, and whether these were meaningful (according to that paper's criteria). Table~\ref{tab:examples} shows a few concrete examples of commit messages, and our classifications for them. 

\begin{table}[htbp]
\caption{Examples of commit messages and our classifications}
\begin{center}
\begin{tabular}{l c c}
\hline
\textbf{\textit{Message}}& \textbf{What} & \textbf{Why}\\
\hline
M1.py update  & $\times$ & $\times$ \\ \hline
added STUFF (justification: xx)  & $\times$ & $\times$ \\ \hline
print 5 results & \checkmark & $\times$ \\ \hline
fix some bug  & $\times$ & \checkmark\\ \hline
fixed indexer & $\times$ & \checkmark \\ \hline
remove test.py (unnecessary file) & \checkmark & \checkmark\\
\hline
\end{tabular}
\label{tab:examples}
\end{center}
\end{table}

The results are shown in Table~\ref{tab:messagetypes}. We found that over 80\% of the commit messages made using the official application and 67.6\% of commits made using the command line did not mention any motivation (the ``why'') behind the changes. Only 17.2\% of the commit messages made using the official application, and 27.6\% using the command line, included both the ``what'' and the ``why'' of the commits. In contrast, 76\% of the commit messages made using the modified application included both ``what" and ``why" information. Only 7.2\% of the commit messages made with the modified application were considered meaningless, which was a significant drop compared to 26.8\% made via official application and 20\% made via command line. 
This data indicates that ``nudging'' the students towards a better form of commit messages is an effective method for them to write useful information about code changes.

\begin{table}
\caption{distribution of commit message types}
\begin{center}
\resizebox{\columnwidth}{!}{\begin{tabular}{c c c c c}
\hline
\textbf{\textit{}}& \textbf{{What \& Why}} & \textbf{{No What}}& \textbf{{No Why}}& \textbf{{Neither}}\\
\hline
Command   & 29 (27.6\%)& 26 (24.8\%)&  85 (67.6\%)& 21 (20.0\%)\\
\hline
Official  & 98 (17.2\%)& 163 (28.5\%)&  463 (81.1\%)& 153 (26.8\%)\\
\hline
Modified  & 95 (76.0\%)& 10 (8\%)&  29 (23.2\%)& 9 (7.2\%)\\
\hline
\end{tabular}}
\label{tab:messagetypes}
\end{center}
\end{table}

Additionally, the majority of commit messages included the ``what" information, regardless of the tool used to create them, whether it be the official application (71.5\%), command line (75.2\%), or the modified version (92\%).  Only 23.2\% of the messages made via the modified application did not include any justification, showing a significant decrease compared to commits made using the original application (67.6\%). Despite this, the results of the study suggest that while students may have an understanding of what should be included in an informative commit message, as evidenced by what the students entered using the modified application, the lack of a required justification in other interfaces makes them cut corners, and ignore the justification for the change. 

\begin{table}
\caption{distribution of commit message lengths}
\begin{center}
\resizebox{\columnwidth}{!}{\begin{tabular}{c c c c c}
\hline
\textbf{\textit{}}& \textbf{{Median/Mean Length}} & \textbf{{ $< 15$ (words)}}& \textbf{{$15-20$ (words)}}& \textbf{{$> 20$ (words)}}\\
\hline
Command  & 5/6 (words)& 99 (94.3\%)&  5 (4.8\%)& 1 (1.0\%)\\
\hline
Official & 4/6 (words)& 543 (95.1\%)&  14 (2.5\%)& 14 (2.5\%)\\
\hline
Modified & 16/20 (words)& 55 (44.0\%)&  30 (24.0\%)& 40 (32.0\%)\\
\hline
\end{tabular}}
\label{tab:length-dist}
\end{center}
\end{table}

Previously, Dyer et al. found that more than two thirds commit messages on open source projects were less than a sentence long~\cite{boa-infrastructure-2013}. For the purpose of this study, we also evaluated the length of the commit messages (Tab.~\ref{tab:length-dist}). The median length of the commit messages made using the original application and command line was similar to each other, both being around four and five words long. About 95\% of the messages made both using the official application and command line were less than one descriptive sentence (less than 15 words), which is even worse than in Dyer et al.'s paper (90\%)~\cite{boa-infrastructure-2013}. In contrast, the median length of commit messages made using the modified application was 16 words long, and 56\% of the messages were at least a normal sentence long (greater than or equal to 15 words). The observations indicate that students tend to write longer and more informative commit messages when they are ``nudged'' to do so.

We performed a two-sample Welch's t-test to compare the mean lengths of commit messages generated via different tools. The results indicate that there are statistically significant differences between the modified application and both the official application (p-value $<< 0.001$) and the command line tool (p-value $<< 0.001$). There was no statistical significant difference between the mean lengths of commit messages made via the official application and the command line tool (p-value = 0.7286). This suggests that the two tools produce commit messages of similar lengths.

Previous studies have suggested that developers tend to write low-quality commit messages due to a lack of motivation and time~\cite{boa-infrastructure-2013,a-good-commit-2022,liu2018neural-far}; our results lead us to speculate that a better UI might help reduce the number of low-quality messages.

\section{Threats to validity}
In classifying the students' commit messages according to the criteria stated in~\cite{a-good-commit-2022}, we encountered many situations that were very difficult to classify -- mostly because the messages were too short. We mitigated this problem by having both authors perform the assessment together, referring to the few examples given in Tian et al.’s paper in order to resolve the conflicts of opinion. While sometimes very short messages are warranted and meaningful, other times they are neither; while length can be a hint for quality, it is not 100\% correlated with quality. Nevertheless, it is possible that other judges might interpret Tian et al.’s criteria differently, leading to different results and conclusions.

\section{Conclusion}
It is essential for developers to understand the importance of writing high-quality commit messages in order to ensure the longevity of a software development project. This work has shown that providing guidance on the components of a good commit message within version control tools can improve the quality of commit messages written by students. This approach requires minimal resources for training and can easily be scaled to larger groups. In the future, it would be valuable to conduct further studies in real-world developer communities to validate these findings.

\bibliographystyle{ieeetr}
\bibliography{main}

\begin{thebibliography}{1}

\bibitem{boa-infrastructure-2013}
R.~Dyer, H.~A. Nguyen, H.~Rajan, and T.~N. Nguyen, ``Boa: A language and
  infrastructure for analyzing ultra-large-scale software repositories,'' in
  {\em 2013 35th International Conference on Software Engineering (ICSE)},
  pp.~422--431, 2013.

\bibitem{a-good-commit-2022}
Y.~Tian, Y.~Zhang, K.-J. Stol, L.~Jiang, and H.~Liu, ``What makes a good commit
  message?,'' in {\em Proceedings of the 44th International Conference on
  Software Engineering}, ICSE '22, (New York, NY, USA), p.~2389–2401,
  Association for Computing Machinery, 2022.

\bibitem{doubt-try-three-2022}
A.~Berg, S.~Osnes, and R.~Glassey, ``If in doubt, try three: Developing better
  version control commit behaviour with first year students,'' in {\em
  Proceedings of the 53rd ACM Technical Symposium on Computer Science Education
  - Volume 1}, SIGCSE 2022, (New York, NY, USA), p.~362–368, Association for
  Computing Machinery, 2022.

\bibitem{liu2018neural-far}
Z.~Liu, X.~Xia, A.~E. Hassan, D.~Lo, Z.~Xing, and X.~Wang,
  ``Neural-machine-translation-based commit message generation: how far are
  we?,'' in {\em Proceedings of the 33rd ACM/IEEE International Conference on
  Automated Software Engineering}, pp.~373--384, 2018.

\end{thebibliography}

\vspace{12pt}

\end{document}